\documentstyle[multicol,prl,aps,epsfig]{revtex}

\begin{document}

\twocolumn[\hsize\textwidth\columnwidth\hsize\csname @twocolumnfalse\endcsname

\title{Scaling behavior of impurities in mesoscopic Luttinger liquids}

\draft
\author {V.\ Meden$^1$, W.\ Metzner$^2$, U.\ Schollw\"ock$^3$, 
and K.\ Sch\"onhammer$^1$}
\address{$^1$Institut f\"ur Theoretische Physik, Universit\"at
  G\"ottingen, Bunsenstr.\ 9, D-37073 G\"ottingen, Germany\\
$^2$Institut f\"ur Theoretische Physik C, RWTH Aachen, D-52056 Aachen,
Germany, \\
$^3$Sektion Physik, Universit\"at M\"unchen, Theresienstr.\ 37,
D-80333 M\"unchen, Germany}

\date{April 18, 2001}
\maketitle

\begin{abstract}
Using a functional renormalization group we compute the flow of the 
renormalized impurity potential for a single impurity in a Luttinger 
liquid over the entire energy range - from the microscopic scale of a 
lattice-fermion model down to the low-energy limit.
The non-perturbative method provides a complete 
real-space picture of the effective impurity potential.   
We confirm the universality of the open chain fixed
point, but it turns out that very large systems ($10^4$-$10^5$ 
sites) are required to reach the fixed point for realistic choices 
of the impurity and interaction parameters.
\end{abstract}


\vskip 2pc]
\vskip 0.1 truein
\narrowtext

The low-energy physics of one-dimensional interacting electron 
systems with Luttinger liquid (LL) behavior is  dramatically affected
by the presence of a single impurity.\cite{LutherPeschel,Mattis,ApelRice,KaneFisher,EggertAffleck,MatveevGlazman}
The problem is usually mapped onto an effective field theory 
using bosonization, where terms which are expected to be irrelevant 
in the low-energy limit are 
neglected.\cite{LutherPeschel,Mattis,ApelRice,KaneFisher,EggertAffleck}
Then forward and backward impurity scattering decouple, and the more 
important backscattering processes are modeled by a single amplitude 
$V_B$.
From a perturbative bosonic renormalization group (RG) calculation 
\cite{KaneFisher} and a boundary conformal field theory analysis
\cite{EggertAffleck} the following picture emerged: 
In a chain of spinless fermions\cite{spinless} with 
repulsive interactions (LL parameter $K_{\rho}<1$) the backscattering
amplitude $V_B$ is a relevant perturbation which grows as 
$\Lambda^{K_{\rho} - 1}$ when the flow parameter $\Lambda$ is sent 
to zero, and the perturbative analysis breaks down. 
This behavior can be traced back to the power-law 
singularity of the $2k_F$ density response function in a 
LL.\cite{LutherPeschel,Mattis} 
On the other hand a weak hopping $t_{w}$ between the open ends of 
two semi-infinite chains is irrelevant and scales to zero as 
$\Lambda^{K^{-1}_{\rho} - 1}$.\cite{KaneFisher} 
{\it Assuming} that the open chain represents the only stable fixed 
point it was argued that at low energy scales and even for a weak 
impurity physical observables behave as if the system was split in 
two chains with open boundary conditions at the end 
points.\cite{KaneFisher}
Here we focus mainly on the local spectral weight $\rho_j(\omega)$ 
for lattice sites $j$ close to the impurity and energies $\omega$ 
close to the chemical potential $\mu$. 
For $\rho_j(\omega)$ a power-law suppression 
$\rho_j(\omega) \sim |\omega|^{\alpha_B}$
with the {\it boundary exponent}\/ $\alpha_B = K_{\rho}^{-1} -1$ 
which only depends on the interaction strength and band filling, but
not on the impurity parameters, was predicted.\cite{KaneFisher} 
Within the bosonic field theory the above conjecture was 
verified by refermionization \cite{KaneFisher}, quantum Monte Carlo 
calculations,\cite{Moon,Reinhold} and the thermodynamic Bethe 
ansatz.\cite{Fendley}

To confirm the field theoretical scenario and the validity of the 
underlying assumptions for a microscopic fermionic 
system with LL behavior,
numerical methods (exact diagonalization [ED], density-matrix 
renormalization group [DMRG]) were applied to the lattice model 
of spinless fermions with nearest neighbor 
interaction.\cite{EggertAffleck,Qin,Rommer,OC}   
Comparing ED data for up to $N=23$ sites with the field 
theoretical prediction for the finite size corrections of energies, 
the expected scaling was confirmed for both weak impurities and weak 
hopping.\cite{EggertAffleck}
However, due to the limited system size it was impossible to go 
beyond the perturbative (in either $V_B$ or $t_{w}$) regime. 
Later it was claimed that the full flow from a weak impurity to the
open boundary fixed point (BFP) was successfully 
demonstrated,\cite{Qin,Rommer} although this strong statement is 
not really supported by the numerical data presented. 
The smallest temperature discussed in Ref.\ \cite{Rommer} corresponds 
to a system of around $300$ lattice sites and the largest system 
considered in Ref.\ \cite{Qin} was $N=52$, 
while in Ref.\ \cite{OC} it was shown that $N \approx 10^2$ 
lattice sites are clearly not enough to exclude an asymptotic
behavior not governed by the BFP, even if one starts out with a 
fairly strong impurity.
 
Recently functional RG methods, originally developed in a field 
theoretical context, 
have been introduced as a new powerful tool in the theory of 
interacting Fermi systems \cite{Salmhofer1}, with applications
so far concentrating on translationally invariant
two-dimensional systems.\cite{2dsystems}
In this letter we apply such a functional RG scheme to the
spinless fermion model with site or hopping impurities.
We compute the complete {\em coupled}\/ flow of the renormalized 
onsite energies and the renormalized hopping amplitudes 
from the microscopic energy scale down to the infrared 
fixed point. 
The flow equations are 
{\em non-perturbative in the impurity strength}\/ while perturbative 
in the electron-electron interaction.
We treat the {\em full functional form}\/ of the renormalized 
impurity potential as generated by the flow, instead of replacing it 
approximately by the scattering amplitudes at the Fermi level.
Computing the local density of states near the impurity we
convincingly confirm the universality of the BFP. However, it
turns out that very large systems ($10^4-10^5$ sites) are required 
to reach the BFP for intermediate impurity and interaction 
parameters.
Our RG scheme is checked against numerical exact DMRG data
for systems with up to $N=768$ sites.

The one-dimensional lattice model of spinless fermions with nearest
neighbor hopping amplitude $t=1$ and nearest neighbor interaction
$U$ is given by
\begin{eqnarray}
H = - \sum_{j} \left( c_j^{\dag} c_{j+1}^{} +
  c_{j+1}^{\dag} c_j^{}  \right)
  + U \sum_{j} n_j n_{j+1} ,
\label{spinlessfermdef}
\end{eqnarray}
in standard second-quantized notation.
Here we focus on the half filled band case.
This model is either complemented by a 
site impurity 
$H_s= V n_{j_0}$ 
or provided with a hopping impurity 
$H_h = -t_{w} (c_{j_0}^{\dag} c_{j_0+1}^{} + \mbox{h.c.})$
on one of its bonds.
 
In a weakly interacting spinless LL with an open end the local 
density of states $\rho_j(\omega)$ near the boundary
can to a surprisingly good approximation be obtained from a 
non-selfconsistent Hartree-Fock (HF) approximation.\cite{boundary} 
It is instructive to consider also the impurity problem within
HF, before turning to the RG treatment.
The impurity leads to Friedel oscillations in the non-interacting 
density profile $\langle n_j \rangle_0$ which for large 
$|j-j_0|$ behaves as $R \sin{\left( 2k_F |j-j_0|\right) }/|j-j_0|$, 
where $R$ is the reflection amplitude of the bare impurity.
Similar oscillations are found in the
matrix element $\langle c_j^{\dag} c_{j+1}^{} \rangle_0$.
Thus both the Hartree potential $U(\langle n_{j-1} \rangle_0 + 
\langle n_{j+1} \rangle_0 )$ and the Fock ``hopping correction'' are
oscillating and of long range. 
One then has to solve a (non-trivial) one-particle problem within 
such a potential and with modulated hopping. 
Taking into account the Hartree term only, the resulting spectral 
weight for $|\omega| \to 0$ shows power-law behavior with an exponent 
which is proportional to the amplitude $UR$ of the 
oscillations.\cite{Klaus}  
We have checked numerically (for systems of up to $10^6$ lattice 
sites\cite{boundary,forthcomming}) that this behavior is not changed 
when the Fock term is included. 
Thus due to the {\it long range}\/ nature of the effective potential 
and the hopping modulation already HF yields a {\it power-law} for 
the spectral weight, but with an exponent which not only depends 
on $U$, but via $R$ also on the {\it bare}\/ impurity strength. 

It is tempting to extend the HF study using self-consistent
HF.\cite{Richter} 
However, it turns out that an iterative solution of the 
selfconsistent HF equation leads for all $U$ to a charge density wave 
groundstate,\cite{forthcomming,Richter} 
which is qualitatively incorrect since a single impurity cannot
change bulk properties of the system.

We now treat the problem using a fermionic functional RG approach. 
Cutting off the free propagator on a scale $\Lambda$ and 
differentiating with respect to this flow parameter, an exact
infinite hierarchy of coupled differential flow equations
for the one-particle irreducible vertex functions can be 
derived.\cite{Wetterich,Morris,Salmhofer2}
For the impurity problem it is technically advantageous to use a 
frequency cut-off for the free propagator
\begin{equation}
 G^{0,\Lambda}(i\omega) = \Theta(|\omega| - \Lambda) 
G^{0}(i\omega)
\end{equation}
where $G^{0}$ is the free propagator without cut-off and $\omega$ 
the Matsubara frequency. $\Lambda$ flows from $\infty$ to $0$.
For spinless fermions the electron-electron interaction
is renormalized only by a finite amount of order
$U^2$.\cite{Solyom} 
Hence, we can replace the renormalized two-particle vertex 
to leading order in $U$ by the antisymmetrized bare interaction.
In this way the exact hierarchy of flow equations gets truncated,
and one obtains a simple one-loop flow equation for the self-energy 
$\Sigma$, where only the (full) propagator $G$ and the bare
electron-electron interaction $U$ enter. 
Carrying out a Matsubara sum and choosing a real space
representation of $G$ and $\Sigma$, one obtains the flow equations
(at temperature $T=0$)
\begin{eqnarray}
\frac{d}{d \Lambda}\Sigma^{\Lambda}_{j,j} & = & - \frac{U}{2\pi} 
  \sum_{s=\pm 1} \sum_{\omega = \pm \Lambda}
  G^{\Lambda}_{j+s,j+s} (i \omega) 
  \label{diffsystem1} \\
\frac{d}{d \Lambda} \Sigma^{\Lambda}_{j,j\pm 1} & = & \frac{U}{2\pi} 
  \sum_{\omega = \pm \Lambda}
  G^{\Lambda}_{j,j \pm 1} (i \omega) \; .
\label{diffsystem2}
\end{eqnarray}  
The self-energy is frequency independent and tridiagonal,
since the bare interaction is instantaneous and restricted to
nearest neighbors.
The full propagator $G^{\Lambda}$ on the right hand side of the
flow equations is obtained by inverting 
the matrix $\left[ G^0 \right]^{-1} - \Sigma^{\Lambda}$.
The bare site and/or hopping impurity enter as 
{\em initial conditions} for $\Sigma^{\Lambda}$ at $\Lambda = \infty$.
For a site impurity $V$ at $j_0$, one sets
$\Sigma^{\Lambda=\infty}_{j_0,j_0} = V$, 
and for a hopping impurity between $j_0$ and $j_0+1$, one has
$\Sigma^{\Lambda=\infty}_{j_0,j_0+1} = 
 \Sigma^{\Lambda=\infty}_{j_0+1,j_0} = -t_{w}$, while the
other matrix elements are initially zero.
The above flow equations are {\it non-perturbative} in the impurity 
parameters, in contrast to the perturbative bosonic RG. 
Written in momentum space the different scattering channels 
$\Sigma_{k,k'}$ are coupled.
The self-energy at $\Lambda=0$ can be given a simple physical 
meaning: 
$\Sigma_{j,j}^{\Lambda=0}$ represents an effective one-particle
potential and $\Sigma_{j,j+1}^{\Lambda=0}$ is an effective modulation
of the hopping. To calculate $\rho_j(\omega)$ one determines 
the spectral weights of the remaining one-particle problem.

For small impurity strength $V$, after transforming  to
momentum space and taking $N\to \infty$, 
Eqs.\ (\ref{diffsystem1}) and (\ref{diffsystem2}) 
can be solved {\it analytically}, as long as $\Sigma^{\Lambda}$ stays
small.  
For the backscattering this gives 
$\Sigma_{k_F,-k_F}^{\Lambda} \sim \Lambda^{-\eta}$ with 
$\eta=U [1-\cos{(2 k_F)}] /(\pi v_F)$ and the Fermi velocity $v_F$. 
To leading order in $U$, the exponent $\eta$ is just
$K_{\rho} -1 $\cite{boundary} which shows that the non-perturbative 
fermionic RG captures the power-law increase found in the 
perturbative bosonic RG. 

\begin{figure}[hbt]
\begin{center}
\vspace{-0.0cm}
\leavevmode
\epsfxsize6.5cm
\epsffile{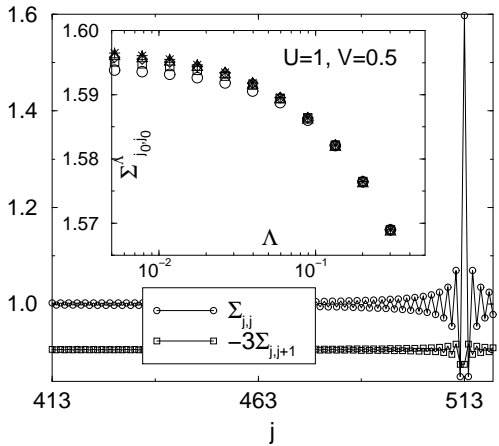}
\caption{$\Sigma_{j,j'}^{\Lambda=0}$ for a {\it site impurity} with 
  $N=1026$ and $j_0=513$. The inset shows $\Sigma_{j_0,j_0}^{\Lambda}$
  as a function of $\Lambda$ for  $N=66$ (circles),
  $N=130$ (squares), $N=258$ (diamonds), $N=514$ (triangles), and
  $N=1026$ (stars).}
\label{figeffpota}
\end{center}
\vspace{-0.0cm}
\end{figure}

Numerically integrating the RG equations for finite systems 
we can go beyond the perturbative regime. In each step of the
integration we have to invert an $N \times N$ matrix. 
If we assume open boundary conditions in $H_0$, 
$\left[  G^{0} \right]^{-1} - \Sigma^{\Lambda}$ is tridiagonal 
in real space and the numerical effort is considerably 
reduced.\cite{note2} 
This allowed us to treat systems with up to 
$2^{15} = 32768$ lattice sites. For finite $N$ the flow
is effectively cut off on a scale of the order of $1/N$.  
For smaller systems we also considered periodic boundary 
conditions.
Fig.\ \ref{figeffpota} shows typical results for 
$\Sigma_{j,j}^{\Lambda=0}$ and $\Sigma_{j,j+1}^{\Lambda=0}$ for a 
{\it site impurity} and lattice sites close to $j_0$. 
Since $\Sigma$ is symmetric around $j_0$ mainly the region $j<j_0$ 
is shown. 
Similar to HF the effective potential and hopping are oscillating 
and slowly decaying. 
The numerical data suggest that the oscillations again fall off 
as $|j-j_0|^{-1}$, but with an amplitude which compared to HF is 
slightly enhanced. 
The inset of Fig.\ \ref{figeffpota} shows 
$\Sigma_{j_0,j_0}^{\Lambda}$ as a function of $\Lambda$ for
different $N$.
Obviously the renormalized potential at the impurity site remains 
finite and the expected ``cutting'' of the chain does certainly 
not occur because a single on-site energy diverges, as one might
guess if the bosonic RG is taken too literally. 
Singular behavior is only found in $\Sigma_{k,k'}^{\Lambda}$ 
for momenta with $k-k' \approx \pm 2k_F$, which is associated
with the {\it long range oscillations} in real space.  

\begin{figure}[thb]
\begin{center}
\vspace{-0.0cm}
\leavevmode
\epsfxsize6.5cm
\epsffile{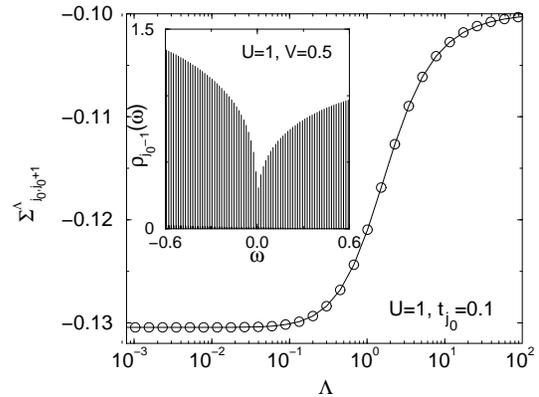}
\caption{ $\Sigma_{j_0,j_0+1}^{\Lambda}$ as a function of $\Lambda$
  for a {\it hopping impurity} with $t_{j_0} = 0.1$, $U=1$, $N=1024$,
  and $j_0 =512$.
  Inset: $\rho_{j_0-1}(\omega)$ as a function of $\omega$
  for a {\it site impurity} with the same parameters as in Fig.\ 1 
  ($N=1026$).}
\label{figdosa}
\end{center}
\vspace{-0.0cm}
\end{figure}

In Fig.\ \ref{figdosa} we present results for 
the case of a {\it weak hopping} between two open chains.  
It shows $\Sigma_{j_0,j_0+1}^{\Lambda}$ as a function of $\Lambda$ for
$N=1024$. We have checked that the curve to a good approximation
already presents the $N \to \infty$ result. 
In contrast to a simplistic interpretation of the bosonization result,    
the renormalized hopping $\Sigma_{j_0,j_0+1}^{\Lambda}$ does {\it not}
scale to zero. 
Similar to the case of a site impurity, $\Sigma^{\Lambda=0}_{j,j'}$ 
shows long range oscillations in both the effective potential 
and the hopping. 
Again this and not the scaling of a single $V_j$ or $t_j$ is the 
reason for the peculiar behavior of physical observables, as 
for example $\rho_{j}(\omega)$, discussed next.

As an inset to Fig.\ 2 the density of states near the impurity, 
$\rho_{j_0-1}(\omega)$, is presented for a {\it site impurity.} 
The  data show a suppression of the weight for $|\omega| \to 0$ 
as expected.
Each spike represents a $\delta$-peak of the finite system.
Instead of trying to fit a power-law to these data 
it is advantageous to analyze the finite size scaling of the 
spectral weight $W(N)$ at $\mu$.\cite{defofw} 
If $\rho_{j_0-1}(\omega)$ follows a power-law as a function 
of frequency, we expect a power-law with the same exponent in 
the $N$ dependence of $W(N)$.  

\begin{figure}[thb]
\begin{center}
\vspace{-0.0cm}
\leavevmode
\epsfxsize6.5cm
\epsffile{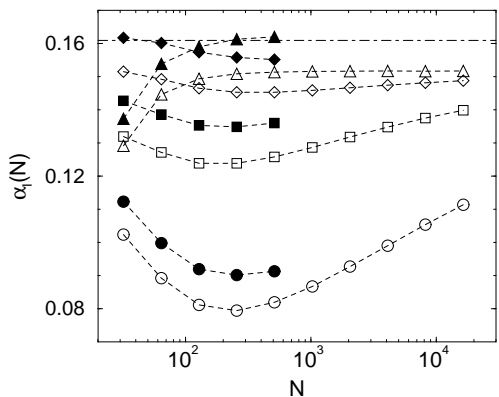}
\caption{$\alpha_I(N)$ as a function of $N$ for $U=0.5$ and different
  $V$: $V=1$ (circles), $V=2$ (squares), $V=4$ (diamonds), and $V=\infty$
  (triangles). The filled symbols are DMRG data and the open ones
  obtained from the RG. The dashed-dotted line gives the exact boundary
  exponent $\alpha_B^{\rm ex}$.}
\label{figUfixed}
\end{center}
\vspace{-0.0cm}
\end{figure}

\begin{figure}[thb]
\begin{center}
\vspace{-0.0cm}
\leavevmode
\epsfxsize6.5cm
\epsffile{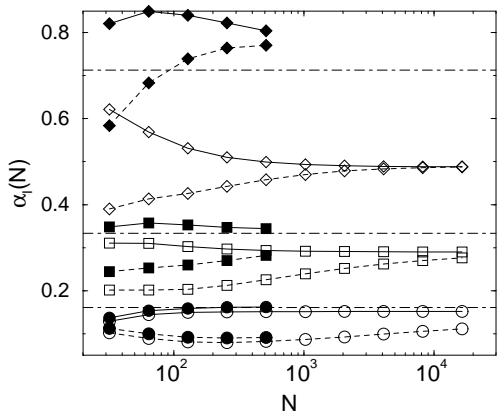}
\caption{$\alpha_I(N)$ as a function of $N$ for $V=1$ (dashed lines)
  and $\alpha_B(N)$ for $V=\infty$ (solid lines) for different 
  $U$: $U=0.5$ (circles), $U=1$
  (squares), and $U=1.8$ (diamonds). 
  Filled symbols are DMRG data, open ones RG results. 
  The dashed-dotted lines give the exact $U$ dependent 
  boundary exponents $\alpha_B^{\rm ex}$.}
\label{figVfixed}
\end{center}
\vspace{-0.0cm}
\end{figure}

In Fig.\ \ref{figUfixed} we show the negative of 
the logarithmic centered differences $\alpha_I(N)$ of 
$W(N)$ as a function of $N$ for $U=0.5$ and different $V$
obtained from RG ($N \leq 32768$) and DMRG ($N \leq 768$).
If $W(N)$ decays for $N \to \infty$ as a power-law, 
$\alpha_I(N)$ converges to the respective exponent. 
For comparison we also calculated $\alpha_B(N)$ for the lattice 
site next to an open boundary ($V=\infty$). 
The DMRG and RG data are parallel to each other, which in addition
to the analytical arguments is a strong indication that our fermionic 
RG captures the essential physics.
For $V=\infty$ both methods produce the expected power-law behavior 
with boundary exponents $\alpha_B^{\rm DMRG}$ and $\alpha_B^{\rm RG}$.
$\alpha_B^{\rm DMRG}(N=512)$ agrees up to $1$\% with the exact
exponent $\alpha_B^{\rm ex} \approx 0.1609$. 
$\alpha_B^{\rm RG}(N=16384)$ which
effectively is equal to $\alpha_B^{\rm RG}(N=\infty)$
deviates by roughly $6$\% from $\alpha_B^{\rm ex}$ 
since the RG is only correct to leading order in $U$.
The RG curves for finite $V$ suggest that for $N \to \infty$ 
the $\alpha_I^{\rm RG}(N)$ converge to the universal 
($V$ independent) exponent $\alpha_B^{\rm RG}$. 
This is in agreement with the field theoretical prediction.
It is remarkable that even for fairly strong impurities 
($V = 4$) extremely large $N = 10^4$-$10^5$ are needed to exclude 
non-universal ($V$ dependent) fixed points with some certainty.
Solely relying on DMRG data for a few hundred lattice sites 
would in this case give no definite result.\cite{OC}
In Fig.\ \ref{figVfixed} RG and DMRG data are presented for an 
intermediate impurity strength $V=1$ and $V=\infty$ for 
different values of $U$. 
Due to higher order corrections in $U$, the
difference between the RG and DMRG data increases with increasing $U$.
For larger $U$, the $\alpha_I^{\rm RG}(N)$ approach 
$\alpha_B^{\rm RG}$ faster, but 
even for the largest $U=1.8$ considered here\cite{limit} 
(which corresponds to $K_{\rho} \approx 0.58$) very large $N$ are needed.
This demonstrates that for intermediate $V$ and $U$, which 
are experimentally most relevant, very large systems are needed 
to observe the universal BFP physics. 
For chains which are not long enough a strong system size dependence 
of experimentally extracted exponents must be expected.

We finally note that in the fermionic RG used in Ref.\ 
\cite{MatveevGlazman} flow equations were set up for a single
parameter only: the transmission amplitude at the Fermi level.
Our functional RG flow however indicates that in the non-perturbative
regime different momentum channels are strongly coupled.
Hence, we believe that it is important to take the whole 
renormalized impurity potential profile into account.
The RG equations used in Ref.\ \cite{MatveevGlazman} can also be 
derived within our formalism, if one makes similar crude  
approximations.\cite{Herbert} Also we did not find signs of an
enhanced spectral weight as predicted in Ref.\ \cite{OregFinkelstein}.

In summary, by solving a functional flow equation
in a fermionic representation we have shown
that in a one-dimensional lattice electron system with 
Luttinger liquid behavior an impurity makes  
observables at low energy scales behave as if the chain 
is split in two parts with open boundary conditions at the end
points. 
Our fermionic RG is non-perturbative in the impurity strength.
Long-range oscillations in the effective impurity
potential provide a simple real-space picture of the ``splitting''
mechanism.
The accuracy of the finite site 
RG scheme was confirmed by a direct comparison to
DMRG data.
For realistic parameters very large systems are needed to 
reach the asymptotic open chain regime.
Hence only special mesoscopic systems, such as very 
long carbon nanotubes, are suitable for experimentally observing the 
impurity-induced asymptotic open boundary physics.
Our method can easily be 
generalized to the case of several impurities and e.g.\ resonance 
phenomena can be studied.\cite{forthcomming}

We would like to thank 
W.\ Apel, P.\ Durganandini, P.\ Kopietz, N.\ Shannon, C.\ Wetterich, 
and especially M.\ Salmhofer and H.\ Schoeller for valuable discussions. 
U.S.\ is grateful to the Deutsche Forschungsgemeinschaft for support
from the Gerhard-Hess-Preis.


\begin{thebibliography}{*}
\bibitem{spinless} Here we will only discuss spinless fermions.
\bibitem{note2} The open ends are far away from the impurity site 
 $j_0$ and do not influence the behavior close to $j_0$.
\bibitem{limit} The LL phase is stable only for $U < 2$.
\end{thebibliography}
\end{document}